# Chemical Liquid Phase Deposition of Thin Aluminum Oxide Films

SUN, Jie*[a](孙捷)    SUN, Ying-Chun[b](孙迎春)

[a]*Key Laboratory of Semiconductor Materials Science, Institute of Semiconductors, Chinese Academy of Sciences, Beijing 100083, China*

[b]*Department of Foreign Languages, Shandong University, Weihai, Shandong 264209, China*

Thin aluminum oxide films were deposited by a new and simple physicochemical method called chemical liquid phase deposition (CLD) on semiconductor materials. Aluminum sulfate with crystallized water and sodium bicarbonate were used as precursors for film growth, and the control of the system's pH value played an important role in this experiment. The growth rate is 12 nm/h with the deposition at $[Al_2(SO_4)_3] = 0.0837$ mol•L$^{-1}$, $[NaHCO_3] = 0.214$ mol•L$^{-1}$, 15 ℃. Post-growth annealing not only densifies and purifies the films, but results in film crystallization as well. Excellent quality of $Al_2O_3$ films in this work is supported by electron dispersion spectroscopy, Fourier transform infrared spectrum, X-ray diffraction spectrum and scanning electron microscopy photograph.

## Introduction

Inorganic oxide films have attracted a lot of interest in the last several decades. Among them, silicon dioxide films are widely used in modern microelectronics, optics and mechanics. This material has been grown by various methods including thermal oxidation, chemical vapor phase deposition, plasma-enhanced chemical vapor phase deposition, and so on.[1,2] Recently, Nagayama *et al.*[3] have reported that $SiO_2$ thin films could be produced by a new chemical method of liquid phase deposition (LPD). It is superior to other deposition methods in terms of the following characteristics: low processing temperature, simple equipment, high growth rate and low cost, and thus has attracted plenty of theoretical and industrial interest.[4-13] The deposition mechanism of LPD is not understood very thoroughly,[14] despite a Raman spectroscopic study of the $H_2SiF_6$ growth solution.[15] This notwithstanding, the technique of LPD has been extended to formation of other oxides, including those of Ti, Sn, Zr, V, Cd, Zn, Ni and Fe.[16-23] To date, however, no liquid phase deposited aluminum oxide thin film has been reported before. In this paper, we will focus on our results of thin $Al_2O_3$ films grown by chemical liquid phase deposition (CLD). The purpose of adding the word "chemical" to "liquid phase deposition" is to emphasize that it is not a physical method of film fabrication, but a chemical one.

Aluminum oxide is an interesting coating material. By virtue of excellent and stable chemical and physical properties, thin $Al_2O_3$ films are extensively applied in optics, solid-state electronics, micro-electromechanical systems (MEMS) and other related spheres. One of its applications is in electroluminescent (EL) structures since it has a low permeability to alkali ions and a high dielectric constant. Another example is to be seen in the utilization of alumina films as insulator layer in metal-oxide-semiconductor (MOS) transistors in integrated circuit industry.[24] Conventional methods to fabricate $Al_2O_3$ films include plasma activated electron beam evaporation,[25] metal-organic chemical vapor phase deposition (MOCVD) using aluminum acetylacetonate as source material,[26] preparation of photoluminescent alumina films by pyrosol process,[27] direct current reactive magnetron sputtering,[28] remote-microwave-plasma-enhanced chemical vapor phase deposition (RMPECVD),[29] *etc.* The authors here will propose a much more convenient and cheaper way of alumina coating by chemical liquid phase deposition. Traditional LPD technique[3-23] requires a fluoro-anion precursor species in the aqueous growth solution, which is quite harmful to the researcher's health. In this paper, we not only originally extend the LPD technique (we call it CLD) to the fabrication of $Al_2O_3$ films, but also prepare film growth solution with poison-free materials. In the following discussion, CLD is proved to be suitable for producing ultrathin alumina films of nanometer scale.

## Experimental

The apparatus used for CLD-$Al_2O_3$ film growth is shown in Figure 1. This system contains a constant-temperature magnetic heating agitator, a polytetrafluoroethylene beaker (with a lid), a water bath and a holder for the wafer. The appended equipment consists of an acidmeter, a 0.22 μm Teflon filter, an ultrasonic cleaner, *etc*. $Al_2(SO_4)_3 \cdot 18H_2O$, $NaHCO_3$ and deionized (DI) $H_2O$ (the resistivity is 18.6 MΩ·cm) were employed to make growth solution. Firstly, 73.25 mL of saturated aluminum sulfate solution ($[H^+] = 10^{0.26}$ mol·$L^{-1}$) was prepared by dissolving 50 g of $Al_2(SO_4)_3 \cdot 18H_2O$ into 50 mL of DI water. The suspended particles are hydrolytes of $Al^{3+}$, and could be filtered out by the Teflon filter. Secondly, 16.115 g of hyperfine powder of bicarbonate of soda was very slowly added to the above-mentioned solution. Multitudinous carbon dioxide bubbles were therein brought about. After the reaction completely ended, we got 74.72 mL of thick aqueous solution with the pH value being 2.89. Thirdly, it was diluted with DI water in the volume ratio of one to eleven until the pH value equaled 3.75. Having been filtered again using the 0.22 μm filter, the solution was now ready for film growth. Summarily, as-produced growth solution, which is a transparent and thin fluid, was made up at $[Al_2(SO_4)_3]=0.0837$ mol·$L^{-1}$, $[NaHCO_3]=0.214$ mol·$L^{-1}$ (final concentration).

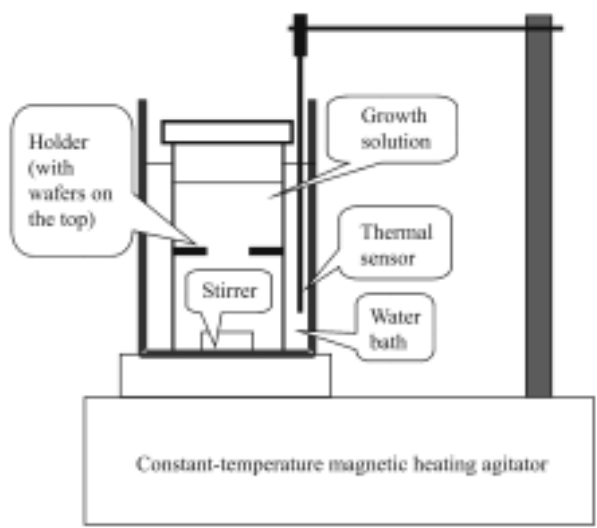

**Figure 1**  Schematic diagram of the apparatus for CLD of $Al_2O_3$ films.

N-type (111)-oriented silicon and semi-insulating (100)-oriented gallium arsenide served as substrate wafers. The Si and GaAs substrates were cleaned by methylbenzene, acetone, and ethanol successively with ultrasonic cleaner in 10 min each. Then, they were etched in a pH-controlled buffer solution (25 ℃, 29% $NH_3 \cdot H_2O$ : $H_2O$ = 1 : 1) adjusted with an acidmeter for 5 min. That etching process is favorable for the growth of oxides.[30] It is requisite for silicon wafers to receive an additional hydrogen peroxide treatment, as will be elucidated in the discussion section of this article. Afterwards, wafers were absterged repeatedly by DI $H_2O$ and then dried up by nitrogen. The Teflon vessel, filled with the newly-prepared $Al_2O_3$ growth solution, a magnetic stirrer and a growth trestle, was immersed in the constant-temperature water bath system. The pretreated semiconductor substrates were submerged in that solution and located onto the polytetrafluoroethylene holder for alumina deposition. The precursor species ($Al_2(SO_4)_3$ and $NaHCO_3$) were hydrolyzed to produce a supersaturated system of aluminum oxide, which then precipitated preferentially on the clean substrate surface, producing a conformal coating. Magnetic stirrer played the function of maintaining the high homogeneity of the growth solution. Controlled by a thermal sensor and an intervalometer, the growth temperature and the deposition time varied from 15 to 35 ℃ and 1 to 20 h respectively. At last, uniform and adherent alumina films were turned out after several hours of high temperature annealing (800—1200 ℃).

## Results and discussion

### Mechanism of film growth

CLD is in essence a transition process of growth solution from a metastable state to a steady one with the system's Gibbs free energy decreasing gradually. That also offers the driving force of film growth. To prepare growth solution is in fact to find the most optimum metastable state. According to the theory of chemical thermodynamics, both solution preparation and film deposition are irreversible processes, so the experimental procedure is as important as the mole percent of growth solution. Demonstrated below is the basic chemical reaction of alumina film growth:

$$\Delta H + Al_2(SO_4)_3 + 6NaHCO_3 = 2Al(OH)_3 + 3Na_2SO_4 + 6CO_2 \quad (\Delta H > 0) \quad (1)$$

That is a very violent hydrolysis procedure of both $Al_2(SO_4)_3$ and $NaHCO_3$. One of the products, $Al(OH)_3$, is an amphoteric hydroxide species. It can be dissolved in acid when pH<3.4, generating $Al^{3+}$; and be dissolved in alkali when pH>12.9, generating $Al(OH)_4^-$. While the pH value is between 4 and 11, by and large, it remains undissolving. Theoretical explanation of that program is not complicated at all and is available in a common inorganic chemistry textbook. Here we provide our experimental results of aluminum hydroxide's dissolubility curve, which is shown in Figure 2. As will be illustrated later, pH control of $Al(OH)_3$'s solvency is the masterstroke during CLD of $Al_2O_3$ films.

Under our conditions, the precursory saturated $Al_2(SO_4)_3$ solution is strongly acidic due to the hydrolysis course of $Al^{3+}$. When sodium bicarbonate solids are compounded into the solution, numerous carbon dioxide bubbles will appear. Nevertheless, sedimentation is not observed since the pH value is still quite small. The reason we choose NaHCO$_3$ powder instead of NaHCO$_3$ solution, obviously, is to maintain the strong acidic circumstances. After the sodium bicarbonate powder is thoroughly dissolved, the solution is altered to the critical transformation point with the pH value being 3.75 by afflux of extrapure water. Turbidity of solution may still occur, however, if the system is not diluted in time. This time the precipitation is not caused by Al(OH)$_3$, but by crystallization of double sulfate (composed of Al$_2$(SO$_4$)$_3$ and Na$_2$SO$_4$). As soon as the pH value of solution reaches 3.75, it is filtered by the 0.22 μm filter and clean semiconductor substrates are placed with the right sides downwards onto the growth trestle (to avoid possible adsorption of large particles). A brief (5 s) ultrasonic treatment will eliminate air bubbles on the wafers. At that very critical point, the solution will be changed from static to dynamic, and dense Al$_2$O$_3$ films generated onto the wafers by the dehydration and polymerization of aluminum hydroxide atomic clusters. Because of the chemical similarity between Al and Ga, the film growth onto GaAs facets can be easily achieved. However, it is quite difficult to fabricate Al$_2$O$_3$ films on Si wafers. We thereof find a way out through producing several monolayers of silicon dioxide by means of soaking the silicon wafers into 30 wt% 60 ℃-H$_2$O$_2$ aqueous solution for 8 h, so that it becomes much easier for Al(OH)$_3$ clusters to form stable chemical bonds with SiO$_2$. Finally, after the procedure of thermal annealing, excellent thin Al$_2$O$_3$ films will be turned out by chemical liquid phase deposition.

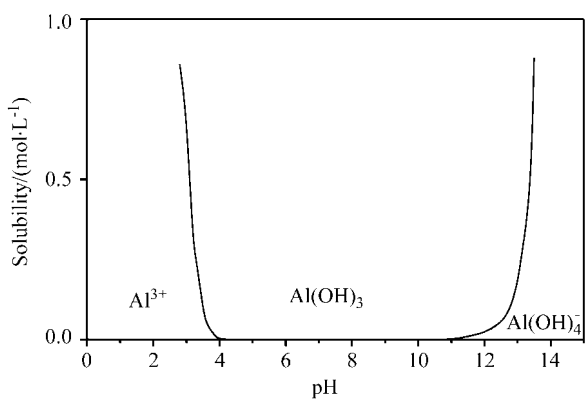

**Figure 2**  Solubility of Al(OH)$_3$ as a function of system's pH value.

**Film growth conditions**

In chemical liquid phase deposition, the growth speed of the films is concentration-controlled. Varying quantities of Al$_2$(SO$_4$)$_3$ and NaHCO$_3$, ranging from 0.06 to 0.10 mol•L$^{-1}$ and from 0.1 to 0.4 mol•L$^{-1}$ respectively, is attentively tested, as is shown in Figure 3. In those domains of concentration, it is learned from Figure 3 that the film deposition rate (the thickness is measured by an ellipsometer) increases at higher NaHCO$_3$ and lower Al$_2$(SO$_4$)$_3$ consistency. Explanation of that effect can be found in Figure 2. All data points in Figure 3 correspond to the region which has a pH value less than 4 in Figure 2. Obviously, data points which have higher NaHCO$_3$ or lower Al$_2$(SO$_4$)$_3$ concentrations are of relatively weak acidity. In other words, they are more adjacent to the Al(OH)$_3$ critical sedimentation point and thereby have enhanced film growth speeds. Comparatively slow growth velocity is not beneficial to a high productivity, but films may be of good quality. That is, the Al$_2$O$_3$ films have less porosity and bigger refractive index. In our experimental circumstances, the optimum growth conditions for the best film we have got are [Al$_2$(SO$_4$)$_3$] = 0.0837 mol•L$^{-1}$, [NaHCO$_3$] = 0.214 mol•L$^{-1}$. At that concentration, we carried out an experimental study of the dependence of film thickness on immersion time at different temperatures. As seen in Figure 4, the thickness of CLD-Al$_2$O$_3$ films increases almost linearly (at the initial stage) with the growth time at assorted temperatures. However, the oxide thickness is seen to saturate after several hours of deposition. That spontaneous termination of film fabrication is attributed to the exhaustion of limited growth sources, namely Al$_2$(SO$_4$)$_3$ and NaHCO$_3$. Eq. (1) is an endothermic chemical reaction, and thus it is reasonable to observe the increased thickness at higher growth temperature in Figure 4. Alumina film's quality can be significantly controlled by adjusting the growth temperature. Too high a temperature e.g. over 40 ℃ will result in poorly adherent and cellular films. To sum up, precursory concentration, submergence time and growth temperature can be used as tools for adapting growth rate, gauge, and certain parameters of final film quality. Additionally, it should also be pointed out that some salts such as aluminum nitrate are not suitable for CLD. The reason is, however, that NO$_3^-$ has a much weaker function of urging Al(OH)$_3$ sol to precipitate than SO$_4^{2-}$ does, and this can be explained by Schulze-Hardy principle.[31]

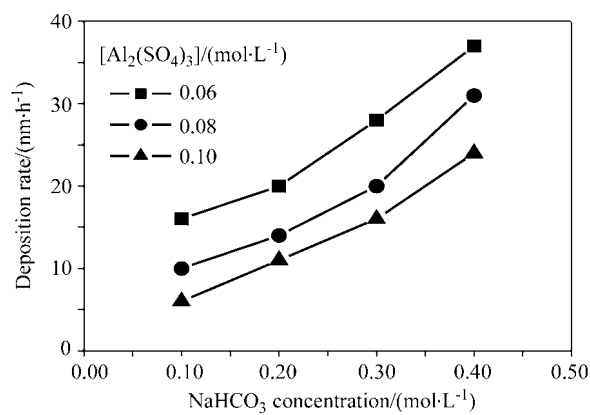

**Figure 3**  Dependence of deposition rate on concentration of precursor species. The samples are 1.5 cm×1.5 cm-Al$_2$O$_3$/GaAs films.

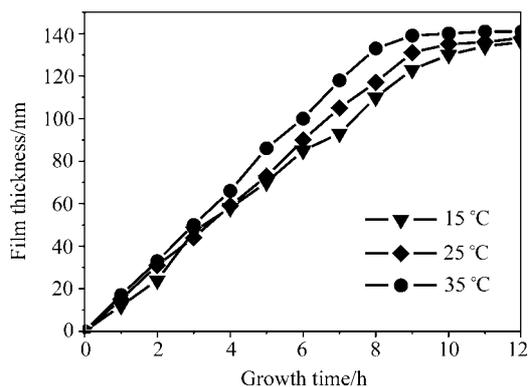

**Figure 4** Dependence of alumina thickness on growth time at different temperatures. The samples are 1.5 cm × 1.5 cm-Al$_2$O$_3$/GaAs films.

### Material characterizations

Displayed in Figure 5 is the electron dispersion spectroscopy (EDS) of an Al$_2$O$_3$/GaAs film deposited at 15 °C for 1 h. It shows clearly four elements contained in the film structure. The inserted figure (see the upper right corner of Figure 5) is the Auger electron spectroscopy (AES) of the same Al$_2$O$_3$/GaAs film. It confirms that the film is composed of Al and O, while the substrate of Ga and As. AES characterization reveals a basically uniform composition of aluminum and oxygen atoms from the surface to oxide-semiconductor interface. Nevertheless, it also reveals a drastic decrease in aluminum content at the interface. That is, the Al/O ratio decreases. In the same zone, an enrichment of Ga in the GaAs is discovered. As can be seen from AES, the interior part of GaAs substrate is arsenide-rich. From all the above data we can draw the conclusion that there is a salient program of Al indiffusion into the GaAs region and Ga outdiffusion from the GaAs surface during oxide deposition. From this conclusion it is strongly suggested that the bond performance of CLD-Al$_2$O$_3$ films is absolutely reliable. Since the Al$_2$O$_3$ films are formed by polymerization of tiny Al(OH)$_3$ particles or clusters which

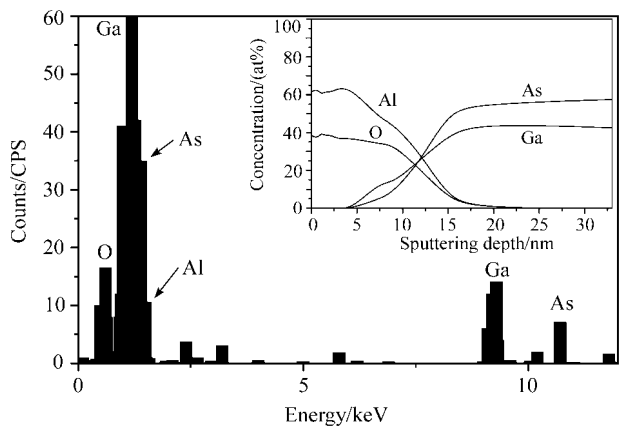

**Figure 5** Energy dispersive X-ray (EDX) spectrum of the CLD-Al$_2$O$_3$/GaAs film. The insertion is AES curve of the same film.

have a large specifical surface area, it is possible for hanging bonds of aluminum and oxygen to remain in the film. That might cause the deviation of atomic ratio between Al and O from 2 : 3. Because chemical liquid phase deposition is a room-temperature procedure, it is likely for hydrogen to exist in the aluminum oxide films. Nevertheless, neither EDS nor AES is sensitive to H so that Fourier transform infrared (FTIR) spectrum is used to detect the existence of hydroxyl groups. On some of the samples, the hydroxyl absorption peak around 3000—3700 cm$^{-1}$ can be seen (not shown in this article). On other samples, hydrogen is not detected and hence negligible. A high temperature treatment is arranged for elimination of H element. Figure 6 is the FTIR spectrum of an Al$_2$O$_3$/GaAs film after furnace annealing for 2 h at 1050 °C. The possibly existing hydrogen is completely removed. Two peaks at 883.28 and 693.15 cm$^{-1}$ are due to the chemical bonds between Al and O, which is consistent with the research of Shek *et al.*[32] Aluminum oxide films will be crystallized into α-Al$_2$O$_3$ (the most steady crystalline phase) after annealing over 1000 °C. In order to study the transition state, we carried out a small incident angle X-ray diffraction (XRD) test of an Al$_2$O$_3$/Si film after two hours of 900 °C-annealing. Due to the exceedingly small thickness value, the diffraction peaks do not appear until the low incident angle is reduced to 0.5°. The curve in Figure 7 indicates that the film's crystalline phase is a

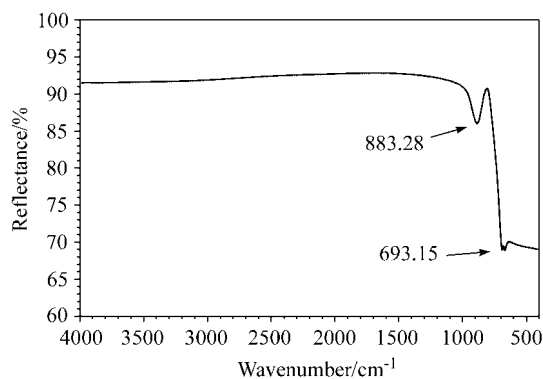

**Figure 6** FTIR spectrum of CLD-Al$_2$O$_3$ film on GaAs substrate after 1050 °C annealing.

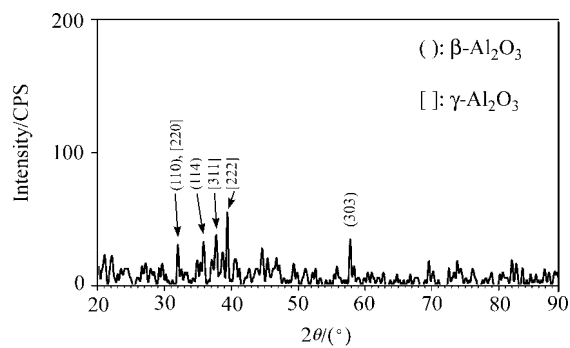

**Figure 7** Small incident angle XRD spectrum of a 900 °C-annealed Al$_2$O$_3$/Si film.

Aluminum oxideChin. J. Chem., 2004, Vol. 22, No. 7

mixture of β-$Al_2O_3$ and γ-$Al_2O_3$ after 900 ℃-annealing. Another proof for high temperature crystallization of the $Al_2O_3$/Si films is offered by a DMH-2LS microhardness tester. The microhardness of the film is 1453.35 kgf/$mm^2$ after 2 h of 1050 ℃-annealing whereas the original hardness value is 520.20 kgf/$mm^2$. The surface morphology of an as-deposited $Al_2O_3$/Si thin film is illustrated in Figure 8. From the scanning electron microscopy (SEM) photograph, we can see that the film is even and uniform, except for a few black mounds caused by the adsorption of large particles during CLD course. More detailedly, a space diagram of the same $Al_2O_3$/Si film surface is observed by atomic force microscopy (AFM), as seen in Figure 9. The AFM data (scan rate is 2.977 Hz) distinctly indicate that the film height undulation in the 3 μm×3 μm square is only 5 nm, and that further assures us a rather excellent exterior.

**Advantages and disadvantages**

A comparison of chemical liquid phase deposition with other $Al_2O_3$ film fabrication methods is summarized in Table 1. Apparently, the growth temperature of CLD is the lowest among all the methods. CLD process is the only one which works at room temperature and pressure. As elucidated before, we employed the furnace annealing program to remove the probably existing hydrogen. However, in the thermogravimetry study, we found that the CLD-$Al_2O_3$ films do not lose weight at a temperature up to 120 ℃, which implies the H ingredient is negligible. That is to say, in most cases the high-temperature annealing process is abridgable, except for a very harsh requirement of the nonexistence of hydrogen. It is known to all that a high-temperature procedure may generate film defects because of large thermal stress, thereby degrading the electronic device characteristics and wiring reliability.[33] Therefore, CLD-$Al_2O_3$ films are supposed to be qualified for electronic applications. Extremely low cost is another chief advantage of CLD over other production means. Raw materials used in CLD are poisonless and cheap. As is displayed in the fifth row of Table 1, the planeness of CLD-$Al_2O_3$ films is comparable or superior to that of films produced by other approaches. A flaw of chemical liquid phase deposition is its extraordinarily low deposition rate. That makes the CLD a time-consuming procedure. In order to fabricate thick $Al_2O_3$ films, it is requisite to soak the wafers in fresh growth solution repeatedly (a re-growth proceeding), for the sources may be depleted each time after roughly 10 h of deposition. That is obviously inconvenient. Conclusively, yielding good quality films, CLD is a room-temperature process with exceedingly low cost, and thereby it offers a fascinating technique of forming $Al_2O_3$ films on semiconductors, especially in the case of ultrathin film applications.

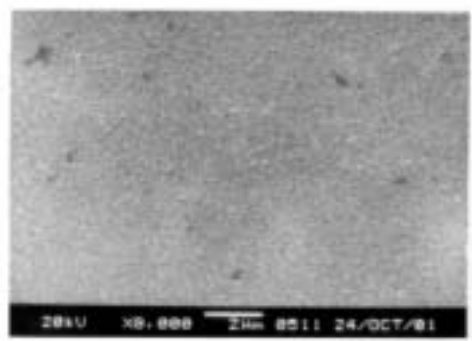

**Figure 8**　SEM micrograph of the CLD-$Al_2O_3$/Si film.

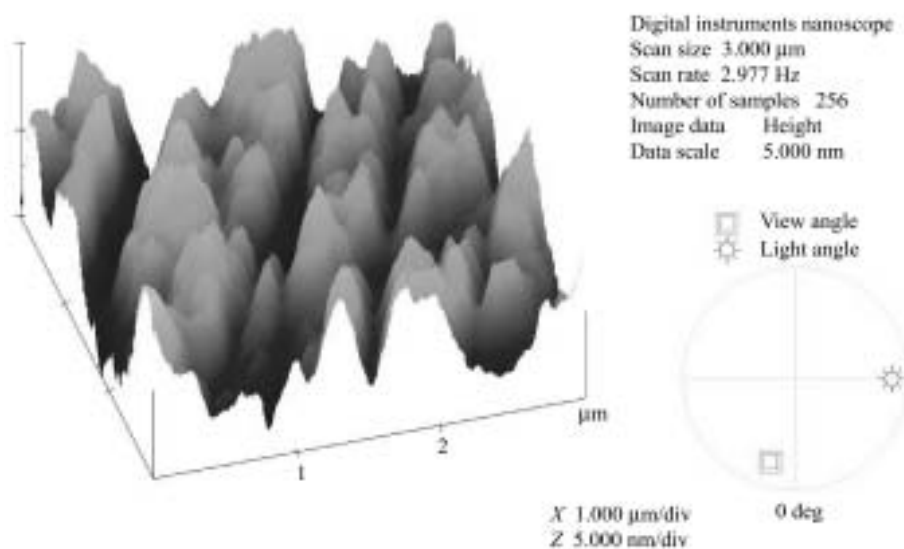

**Figure 9**　AFM stereograph of the CLD-$Al_2O_3$/Si film.

**Table 1**  An outline of characteristics of CLD-$Al_2O_3$ films compared with those by other approaches.

| | Plasma activated electron beam evaporation | Low-pressure metalorganic chemical vapor phase deposition | Pyrosol process | Reactive magnetron sputtering | Remote-microwave-plasma-enhanced chemical vapor phase deposition | Chemical liquid phase deposition |
|---|---|---|---|---|---|---|
| Growth temperature | 100—700 ℃ | 230 ℃ | 480—510 ℃ | <400 ℃ | 540 ℃ | 15—35 ℃ |
| Raw materials | Aluminum, argon, oxygen | Aluminum acetylacetonate, water vapor | Aluminum acetylacetonate, water, methanol | Aluminum, argon, oxygen | Trimethylaluminum, argon, oxygen | Aluminum sulfate, sodium bicarbonate, water |
| Expenditure | High | High | Middle | High | High | Low |
| Surface height undulation | May be of grained structure | — | — | 0.72—2.64 nm within 4.5 $\mu m^2$ | 65—150 nm within 5 $\mu m^2$ | 5 nm within 3 $\mu m^2$ |
| Deposition rate | (0.9—1.8)×$10^5$ nm/h | 150 nm/h | 1.02×$10^3$ nm/h | 1.29×$10^4$ nm/h | 3.6—6.6 mg/($cm^2 \cdot h$) (weight gain) | 12—17 nm/h |
| Reference | 25 | 26 | 27 | 28 | 29 | This paper |

## Conclusions

In summary, we have achieved chemical liquid phase deposited aluminum oxide thin films on semiconductor materials (GaAs and Si). The new technique of CLD and the liquid phase deposition (LPD) method[3] come down in one continuous line. The growth mechanism was investigated and a model for the pH-controlled deposition of $Al_2O_3$ proposed. The growth speed of CLD-$Al_2O_3$ films depends on experimental conditions. Optimized quality was obtained by proper $Al_2(SO_4)_3$ and $NaHCO_3$ concentration ($[Al_2(SO_4)_3] = 0.0837$ mol·$L^{-1}$, $[NaHCO_3] = 0.214$ mol·$L^{-1}$) and comparatively low temperature (15—35 ℃). Various analytical methods like EDS, FTIR, XRD, SEM, *etc*, all show the reliable and good quality of aluminum oxide films made by CLD. Compared with traditional film formation process, CLD is proved to be a candidate for low cost, convenient deposition of alumina on semiconductor substrates. Nevertheless, much still remains to be done to give impetus to chemical liquid phase deposition towards a much more mature technology.

## Acknowledgements

The authors wish to thank Academician Wang Zhanguo and Professor Hu Lizhong for their theoretical directions.